Original Paper

# Topic and Sentiment Trends in Semaglutide Discussions on X: Subpopulation-Based Longitudinal Analysis

Parisa Momeni[1*], M.S.C.S; Gabriel Laverghetta[1*], M.S.C.S; Jay Ligatti[1], PhD; Lingyao Li[2], PhD

[1]Deparment of Computer Science and Engineering, University of South Florida, Tampa, FL USA
[2]School of Information, University of South Florida, Tampa FL USA
* The first two authors contributed equally to this work

**Corresponding Author:**
Parisa Momeni
University of South Florida
Tampa, FL, USA
Email: parisamomeni@usf.edu

## Abstract

**Background:** User experience has a significant impact on pharmaceutical drug effectiveness. Social media platforms like X (formerly Twitter) have become prominent spaces where individuals share their medication-related experiences, especially with widely marketed drugs such as semaglutide. Despite the large volume of conversation, a comprehensive understanding of how various user subpopulations engage with semaglutide-related discussions remain underdeveloped.

**Objective:** This study aims to explore how semaglutide is perceived and discussed across different X user groups. Within these user groups, we investigate (1) the evolution of sentiment patterns toward semaglutide and (2) the evolution and prevalence of semaglutide-related discussion topics.

**Methods:** We prepare a dataset consisting of 859,751 X posts (tweets) pertaining to semaglutide, along with related metadata, that were posted between July 2021 and April 2024. We apply sentiment analysis and topic modeling to the collected posts and analyze the sentiment patterns and topics within specific user subpopulations and time periods.

**Results:** Our analysis reveals a mean sentiment score of -0.242 across all posts, with all user subpopulations experiencing a decline in sentiment during the study period. User discussions focus on semaglutide's applications in weight loss and potential side effects, along with economic factors and celebrity/political influence. We also uncover differences in sentiment and discussion topics across user subpopulations. Notably, organizational accounts consistently express less negative sentiment (mean = −0.04) than individuals (mean = −0.28), with a statistically significant difference ($P < .001$), particularly in discussions related to drug efficacy and regulatory concerns. Interrupted time-series analysis shows a marked decrease in sentiment during the November 2022–January 2023 period, coinciding with regulatory announcements about potential adverse effects. In addition, we observe gender-based variations, such as a greater prevalence of discussions involving celebrities and politicians within female user posts (21%) compared to male user posts (17%) and male users expressing more positive sentiment.

**Conclusions:** This study helps advance the understanding of how diverse user groups perceive and discuss widely marketed drugs like semaglutide. Although we observe a general negativity, there are nuanced differences among the subpopulations. Our results offer valuable implications for health communication strategies and pharmacovigilance.

## Introduction

Semaglutide, also known by brand names such as Ozempic and Wegovy, has surged in popularity in recent years [1]. Originally developed as a diabetes medication, semaglutide has recently shown effectiveness in off-label use for weight-loss treatment [2]. Semaglutide was the fourth highest drug expenditure in the United States in 2021, with $10.8 billion spent on the drug [3]. Social media platforms like X (formerly Twitter) have become key venues for the public to share experiences and express opinions about semaglutide [4]. Celebrities or public figures, such as Elon Musk, have shared personal weight-loss stories and endorsed the drug, further amplifying conversations [1, 5]. The widespread advertising, endorsements by high-profile figures, and increased consumer interest have made semaglutide a trending topic in medications [6]. As such, social media offers a unique lens to examine how it is perceived and discussed, shedding light on public sentiment, misconceptions, and concerns [4, 7].

Understanding public perceptions is essential, as user experiences significantly influence the evaluation of pharmaceutical drugs' effectiveness [8, 9]. Positive experiences not only enhance user satisfaction but also contribute to improved adherence and overall well-being [10]. Mining social media data allows policymakers and pharmaceutical providers to tap into a vast repository of real-time data pertaining to user experiences [11]. In addition, by analyzing user-generated content, researchers can uncover nuanced insights into the concerns, preferences, and challenges faced by specific subpopulations by gender, location, or other demographic attributes. This granular analysis provides an opportunity to identify unmet needs, tailor interventions, and ensure more equitable healthcare outcomes.

In addition to obtaining information from social media, recent research has focused on applying natural language processing (NLP) techniques due to their ability to quickly process large-scale and unstructured information [12, 13]. NLP has been leveraged to extract chemical-disease relations [14], build health knowledge graphs [15], and ease the process of documentation in electronic health records [16] (which often use unstructured and non-standardized formats) [17]. Text analytics approaches such as named entity recognition, topic modeling, and sentiment analysis have been applied within the health context [18, 19, 20].

Although mining drug-related user experiences on social media has been widely explored [21, 22], few studies have focused specifically on semaglutide-related discourse. Our study combines large-scale sentiment analysis and topic modeling with user subgroup analysis, offering a granular view of public engagement with semaglutide. Prior work has primarily leveraged social media to identify adverse reactions to semaglutide that were not detected during clinical trials [7, 23]. The study by Alvarez-Mon [4] includes a manual analysis of 2,045 posts to determine user interests, beliefs, and experiences pertaining to semaglutide and other anti-obesity drugs. However, the public discourse including the sentiments and prevalent topics within specific user groups has been under explored. To uncover patterns in how different user subpopulations experience and discuss semaglutide on X, we investigate the following research questions.

- **RQ1 (Sentiment Analysis)**: What underlying factors explain how sentiment toward semaglutide evolves over time and across different user subpopulations, and what insights can be drawn about public concerns and motivations?
- **RQ2 (Topic Modeling)**: Which topics of discussion are most prevalent in positive and negative semaglutide-related posts across different user subpopulations, how do their prevalence patterns change over time, and what do these patterns reveal about group-specific attitudes, priorities, and health communication needs?

The first research question aims to explore differences in engagement patterns and sentiment expressions across user subpopulations and overtime. The second question aims to identify the various discussion topics emphasized by distinct user subpopulations, exploring the prevalence of these topics among the subpopulations. Addressing these two research questions, our study provides a comprehensive discourse analysis across user subgroups, along with an identification of external events that influence the evolution of these patterns. This insight into the real-world user experience is crucial for tailoring public health communication and improving medicine support strategies for diverse communities.

## Related Work

### Exploring User Experiences via Social Media

Crowdsourcing, originally defined as the act of a company or institution taking a function once performed by employees and outsourcing it to an undefined (and generally large) network of people in the form of an open call, has revolutionized how researchers gather and analyze public opinion [24]. Similar methodological approaches have been applied in other domains to understand public perceptions of urban accessibility and inclusion through crowdsourced online reviews [25]. While conventional methods like polls and surveys remain valuable, crowdsourcing through social media platforms enables researchers to collect and analyze large-scale, near-real-time data about user experiences and perspectives [26].

In the healthcare domain, crowdsourcing via social media has become particularly valuable for understanding public opinions about medical treatments and pharmaceutical drugs [27, 28]. First, social media data has shown invaluable potential in pharmacovigilance, due to the tendency of users to share their opinions or experiences such as adverse drug reactions [29]. Researchers have leveraged social media data to track topic trends [30], estimate disease prevalence [31], and analyze public response to health policies [32]. Second, crowdsourcing has proven effectiveness in capturing user experiences that might not be readily available through traditional clinical studies or surveys [33]. In particular, social media platforms provide researchers with access to diverse user populations and their real-world experiences with pharmaceutical drugs, such as off-label use [9] and adverse reactions [34].

### Analyzing User Experiences via NLP Techniques

Our work makes use of sentiment analysis and topic modeling to uncover patterns in user experiences with semaglutide. Sentiment analysis, also called opinion mining, is a branch of NLP that focuses on classifying people's opinions into positive, negative, or neutral associated with data [35]. Research in this domain spans various levels of granularity, from assigning a single sentiment to an entire document or individual sentences to analyzing distinct aspects linked to specific entities [36]. After the COVID-19 pandemic, there has been increasing interest in using sentiment analysis to evaluate the attitudes, perceptions, and emotions expressed by social media users [37-39]. Numerous studies have focused on platforms such as X, Reddit, and Facebook, which have become prominent spaces for sharing public opinions related to COVID-19 [40].

Topic modeling techniques, such as LDA [41] and BERTopic [42], seek to discover the key themes present in a corpus of documents [43]. Topic models can summarize large datasets by capturing the topics (i.e., the major discourse) that appear most commonly in the text. Numerous works have employed topic models to study health-related discussions on social media. For example, Asghari [44] identifies trending topics pertaining to healthcare on X. Topic analysis has provided insights into news reports surrounding COVID-19 [45] and public opinion regarding blood donation [46]. Another study trains an aspect-based topic model to characterize the health topics and then estimates the prevalence of influenza and allergies over time by observing the number of mentions of each topic during different time periods [47]. In addition, prior work has analyzed semaglutide-related discussions on Reddit via topic modeling [48-50]. Our study focuses on exploring how public discourse and sentiment differ across user subpopulations. Analyzing specific subpopulations enables a nuanced understanding of user concerns, such as accessibility, side effects, and insurance coverage, guiding targeted strategies for addressing subpopulation-specific needs.

## Methods

Figure 1 provides an overview of our research design. We compile a dataset consisting of semaglutide-related X posts (Stage 1) and perform sentiment analysis and topic modeling on these posts (Stage 2). We analyze the results to address our two research questions (Stage 3).

### Data Preparation and User Attributes

We use Brandwatch [51], a social media analytics platform, to collect data from X. Brandwatch uses the X API to obtain posts from prior periods, offering a representative sample of X's entire dataset. Our data collection targets X posts posted between July 1, 2021— when the FDA approved semaglutide for chronic weight management [52]—and April 30, 2024. This nearly three-year timeframe enables our longitudinal discourse analysis surrounding semaglutide within different user communities.

**Figure 1**. Overview of the designed framework to implement the research.

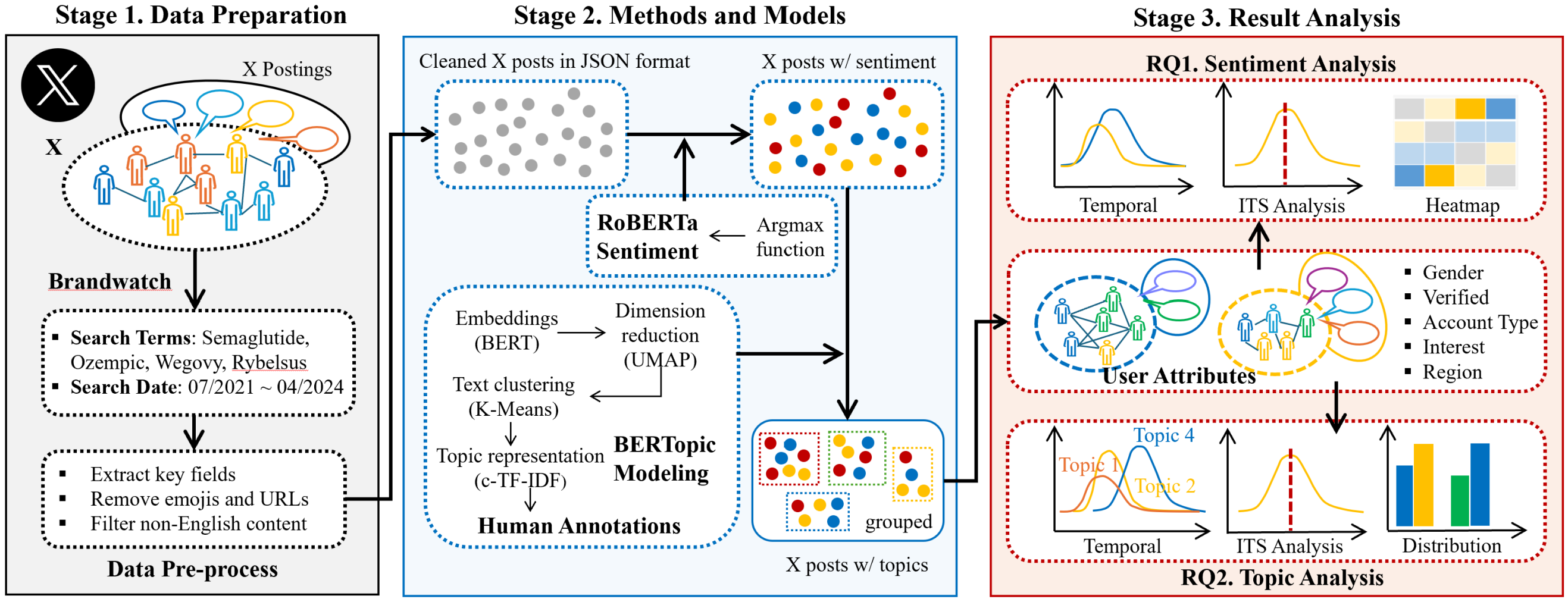

To collect the data, we use several key search terms, including "semaglutide" and its branded names "Ozempic," "Wegovy," and "Rybelsus." We choose these search terms to ensure comprehensive coverage of discussions related to semaglutide and its marketed variants. To maintain consistency, we limit the dataset to English-language X posts containing these terms. The final dataset consists of 859,751 posts, including original posts, replies, reposts (retweets), and quotes.

The user attributes we study are gender, US region, interests, account type, and verification status. These attributes are inferred by the data provider, Brandwatch. According to their documentation, the methodology for this inference varies by attribute; for instance, location is primarily determined from explicit, user-provided information in their X public profiles, whereas attributes like gender and interests are classified using machine learning models that analyze public data such as first names and biographical text [56]. We acknowledge that the specific algorithms used for this demographic inference are proprietary to Brandwatch, and as such, the details of their methods are not publicly available [56].

We choose these attributes for their relevance in capturing diverse user perspectives and behaviors. For example, gender and region can reveal variations in healthcare access and cultural attitudes [53, 54] and have been identified as factors associated with semaglutide initiation [55]. Account type and verification status help differentiate individual users from organizations. Interests reflect personal activities. It should be noted that users may have multiple interests or none, placing them in zero or more interest-based subpopulations.

Since individual users may post multiple posts, we measure the size of each subpopulation by number of users and by number of posts. When grouping each post by its respective user, we assign aggregated values for attributes (gender, region, interests, account type, and verification) using the most frequently observed values across that user's posts. This aggregation ensures that each user is represented by a single record while preserving their attributes. The dataset contains 436,551 unique users. Table S1 and Table S2 in Multimedia Appendix 1 summarize the user subpopulations.

## Ethical Considerations

The University of South Florida Institutional Review Board reviewed this study and determined it to be exempt (IRB ID STUDY009222). To remain compliant with X's policies, we adhere to all API rate limits and do not collect deleted posts. All the code used to generate our results is available from our GitHub repository [57]. To protect user privacy, our full dataset containing identifiers such as usernames is not available for public access [58]. An anonymized version can be accessed from Multimedia Appendix 1 or requested from the corresponding authors. Lastly, we follow standard guidelines [59] to mitigate potential harms from sensitive content contained in our dataset (e.g., suicide mentions).

## RoBERTa Sentiment Analysis

To classify the sentiment of X posts, we utilize the *cardiffnlp/twitter-roberta-base-sentiment-latest*

model from Hugging Face [60], which is pre-trained on X data and widely recognized for its state-of-the-art performance in sentiment classification tasks [61], particularly in handling X posts [62]. We apply the RoBERTa model to each post, obtaining a set of values representing the likelihood of the post having negative, neutral, or positive sentiment. The model outputs these values in a list format, with the elements at indices 0, 1, and 2 corresponding to the negative, neutral, and positive probabilities, respectively [60]. The final sentiment label is determined by the *argmax* function; no additional calibration or thresholding is performed on the sentiment scores. After that, we assign each post a sentiment label, i.e., -1 for negative, 0 for neutral, and 1 for positive. As a result, of the 859,751 total posts, 116,091, 429,074, and 314,586 are classified as positive, neutral, and negative sentiment, respectively.

We calculate mean sentiment scores within each of the user subpopulations (gender, US region, account type, verification status, and interests). To measure sentiment per user, we first group each post by its respective user. We then calculate the mean sentiment score for each user by averaging their per-post sentiment scores, resulting in a continuous value between -1 and 1. This grouping allows us to analyze the aggregated averages of user-level sentiments across the subpopulations. Since the volume of posts varies by subpopulation (e.g., individual user accounts create 1.9 posts on average, compared to 4.4 from organizational users), we also calculate average sentiment per post. In addition, to assess the robustness of our sentiment findings, we conduct a repost-excluding sensitivity analysis, in which we measure average sentiment within each subpopulation while excluding reposts (resulting in 411,747 posts for sensitivity check).

## Interrupted Times-Series Regression

To examine how sentiment evolves over time, we group the dataset and calculate the average sentiment bimonthly spanning from July 2021 to April 2024. This procedure yields a time series of average sentiment scores. Our longitudinal analysis is performed per post as a user can publish multiple posts at different times. To determine whether the observed sentiment shifts exceed baseline trends, we apply an Interrupted Times-Series (ITS) regression. Using bimonthly sentiment averages, we model (1) the baseline time trend, (2) an immediate level change at the intervention point, and (3) a slope change after the event. This approach allows us to separate long-term temporal patterns from abrupt discontinuities. Given the large dataset and multiple comparisons, we treat results as exploratory and emphasize effect sizes and confidence intervals over strict hypothesis testing. Our ITS regression is modeled as follows:

$$Y_t = \beta_0 + \beta_1 Time_t + \beta_2 Event_t + \beta_3 (Time_t Event_t) + \epsilon$$

where $Y_t$ represents the mean sentiment score during the bimonthly period, $Time_t$ is the continuous time index, $Event_t$ is a binary indicator coded 0 before the intervention and 1 afterward, and $Time_t Event_t$ captures the post-event slope. Here, $\beta_0$ estimates the baseline level at the beginning of the series, $\beta_1$ captures the pre-event trend, $\beta_2$ reflects the immediate level change at the intervention, and $\beta_3$ represents the post-event slope change.

## BERTopic Modeling

After performing sentiment analysis, we use the BERTopic model [42] to discover the commonly discussed topics in the dataset. We divide the dataset into posts with a positive RoBERTa sentiment label and posts with a negative sentiment label; posts with neutral sentiment are excluded to focus on identifying the topics that contribute to positivity and negativity. We perform topic modeling for the positive and negative tweets separately. To create the positive and negative document corpora from our dataset, we first clean the text of each post. This cleaning process involves steps such as removing emojis, punctuation, and stop words, normalizing whitespace, and converting all text to lowercase. Cleaning the text is an important step due to the unstructured nature of social media posts [63]. Note that these cleaning steps are not performed prior to obtaining the sentiment of each post, as they may have affected the sentiment results. For example, emojis [64] and punctuation [65] can impact sentiment scores. Our topic modeling exercise focuses solely on the themes present in the text, as opposed to the sentiment of the text. After cleaning each post's text, we remove duplicated reposts to avoid situations in which distinct reposts of a given post are mapped to different topics. The positive and negative document corpora are lists consisting of the snippets of each of the positive and negative cleaned X posts, respectively.

After performing the cleaning steps, we initialize a BERTopic model with default values for all hyperparameters. We run the BERTopic model on our document corpora to generate a list of topics and then extract the sentence/document embeddings for those topics. We use K-means clustering [66] with the Elbow Method to determine the optimal number of topics. Figure S1 and Figure S2 (see Multimedia Appendix 1) show the K-means clustering results for the positive and negative document corpora, respectively. We then run the BERTopic model on our document corpora using 100 clusters.

BERTopic outputs a topic representation and a document representation. In our case, the documents are cleaned X posts. For each topic, the topic representation lists its representative keywords, representative documents, and document count. The BERTopic document representation maps each document in the corpus to its topic number. After reviewing the 100 positive and 100 negative topics produced by BERTopic, we observe that many of the topics shared similar topics. We therefore manually annotate the 200 topics into ten umbrella topics. These topic groupings consist of all the topics that share a common theme. For instance, umbrella topic 0 consists of topics relevant to weight loss.

We perform the manual annotation by reviewing the representative keywords and documents. For example, one of most common positive topics is represented by the following list of keywords: [*semaglutide, semaglutides, weightloss, diet, appetite, medication, eat, treatment, craving, fda*]. A representative document (X post after cleaning) for this topic is: "*ready lose weight gain confidence say hello semaglutide gamechanging prescription medication help achieve significant weight loss with semaglutide take control craving appetite finally reach weight loss goal*". These keywords and the document are associated with themes of weight loss, so we map this topic to umbrella topic 0 (weight loss). As another example, one of the most common negative topics is represented by these keywords: [*nausea, diarrhea, vomiting, nauseous, constipation, vomit, nauseate, constipate, diarrhoea, stomach*]. One of the representative documents for this topic is: "*nausea diarrhea stomach abdominal pain vomiting constipation side affect ozempic*". This topic pertains to adverse reactions experienced after taking semaglutide; therefore, we map it to umbrella topic 8 (acute harm/adverse drug reactions). The mapping of all 200 initial topics to the ten umbrella topics is available in Multimedia Appendix 2. Additionally, to verify that the umbrella topic shares do not change significantly under different cluster counts, we conduct stability checks using 50 and 25 clusters. The results are described in the Umbrella Topic Stability Checks section in Multimedia Appendix 1.

Two of the authors independently map the initial 200 topics to the ten umbrella topics. We calculate the inter-coder agreement between the annotators using Krippendorff's alpha [67], as described by the following equation, where $D_0$ is the observed disagreement between the annotators and $D_e$ is the expected random disagreement:

$$\alpha = 1 - \frac{D_0}{D_e}$$

The inter-coder agreement is 0.806, indicating a satisfactory level of agreement between the annotators. Each X post is mapped to one of the clustered 200 topics using BERTopic, and each of the clustered 200 topics is mapped to one of the ten umbrella topics based on manual annotation. We can therefore map each post to its umbrella topic. Table 3 shows the ten umbrella topics, examples of their representative keywords, and the number of posts mapped to each topic. In addition, Table S3 (available in Multimedia Appendix 1) provides example representative documents (posts) for the ten umbrella topics. During the annotation, the topics are mapped to umbrella topic T9 ("other") if the topic does not clearly match any of the other nine umbrella topics. We observe various sub-topics that appear within T9, such as drug marketing/news, other non-semaglutide drugs, and health/beauty. With the labeled umbrella topics, we group the data by user attributes to discover the prevalence of each umbrella topic among user subpopulations.

## Results

### RQ1 (Sentiment Across User Subpopulations)

Addressing RQ1, we find that the overall sentiment towards semaglutide during the study period (July 2021 to April 2024) is slightly negative, with a mean sentiment score of -0.24 across all posts and –0.28 across all users. Within all user subpopulations, we observe a decline in sentiment over time, but the trend varies across different subpopulations, as discussed in the following subsections.

**Figure 2.** Time-series plots showing bimonthly sentiment analysis categorized by gender, interest, verified status, account type, and region. To improve readability, we limit the interest visualization to the top 5 most popular of the 21 user interests. Shaded areas represent 95% CIs.

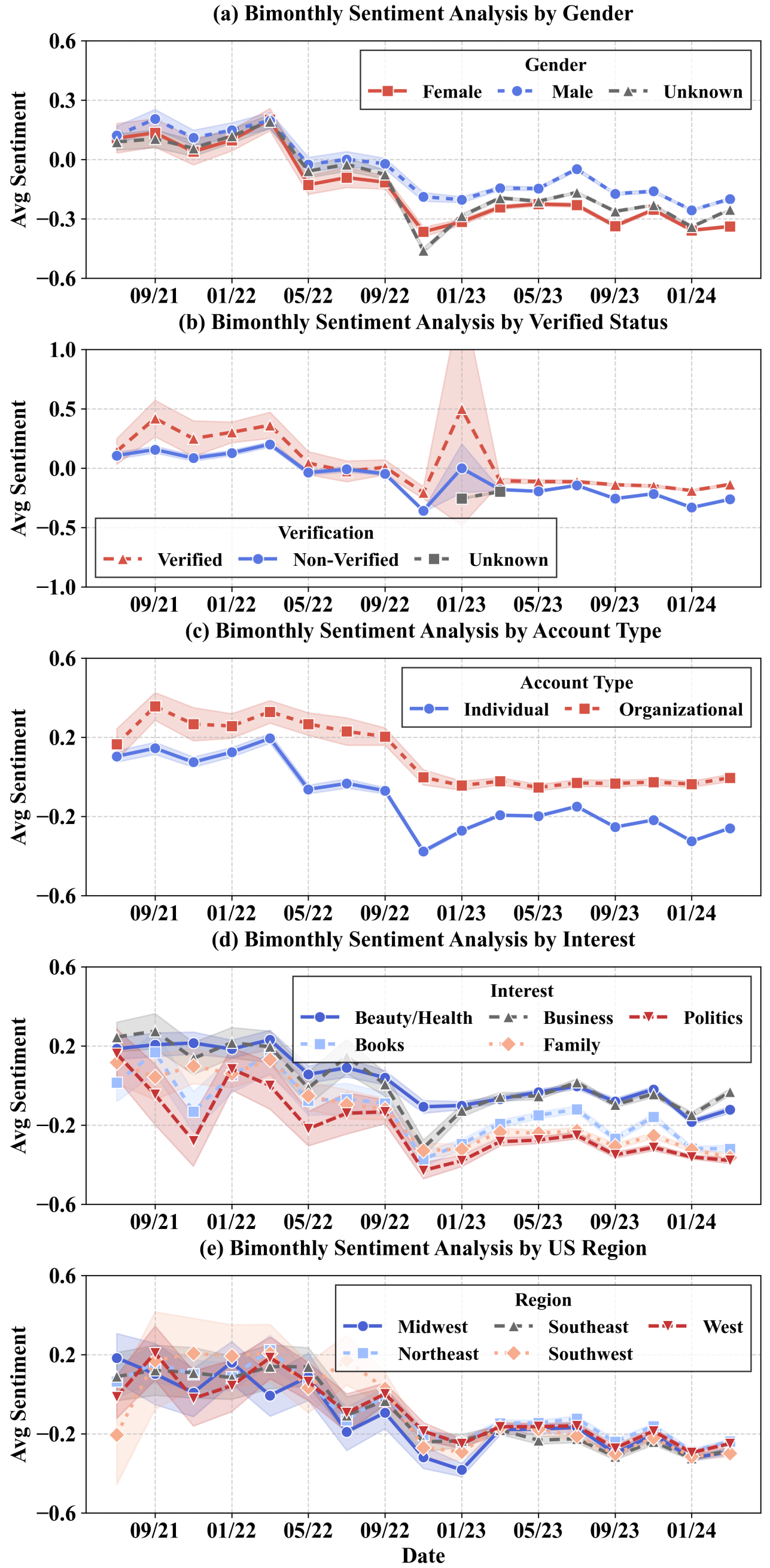

### *Overall Sentiment Declines*

Our longitudinal sentiment analysis results are displayed in **Figure 2**. To illustrate uncertainty, we include 95% confidence intervals (CIs) for key categorical attributes as shaded bands. The temporal progression of sentiment can be divided into four phases: (i) initial positive sentiment across categories (2021- mid 2022); (ii) universal decline (November 2022-January 2023); (iii) variable recovery rates through 2023; and (iv) eventual stabilization at slightly negative levels by early 2024. Three notable periods of universal decline are present: the largest from November 2022-January 2023, the second from September-November 2023, and the final from January-March 2024. Our repost-excluding sensitivity analysis yields similar trends, supporting the robustness of these temporal trends.

### *Connection with External Events*

The sentiment changes are associated with external events (some key events are highlighted in Figure 3). The first clear sentiment decline starting in mid-2022 coincides with the national shortage of GLP-1 medications, which led to increased reliance on compounded semaglutide formulations and raised concerns about access, safety, and regulatory oversight [68]. During this same period, the FDA approved Wegovy for adolescent patients, intensifying public debate around expanded clinical use [69]. Moreover, there is a pronounced decline in sentiment within all user subpopulations during November 2022–January 2023 which coincides with reports of adverse gastrointestinal reaction, suggesting a public concern about potential side effects [70]. Following this decline, a temporary spike in positive sentiment, particularly among verified users, emerges between January 2023–March 2023. However, the wide CI for verified users indicates a high standard error likely resulting from the limited number of verified-user posts within that bimonthly window. This increase in sentiment coincides with heightened media attention and celebrity endorsements of semaglutide as a "Hollywood weight loss drug" [71], suggesting that promotional activity and influencer-driven narratives could temporarily reverse prevailing sentiment patterns.

Following the second dip between September–November 2023, sentiment increases across all user attributes. This recovery aligns with the release of positive cardiovascular outcomes from the SELECT trial, which was presented at the American Heart Association Scientific Sessions and published in the New England Journal of Medicine in November 2023 [72]. The results show that Wegovy (semaglutide 2.4 mg) can significantly reduce major cardiovascular events in adults with overweight or obesity and established cardiovascular disease, potentially improving public perceptions of the drug's efficacy and safety. Following the third dip in sentiment between January and March 2024, we observe a rebound across nearly all subpopulations. This recovery coincides with two major announcements in early March: (i) a NIH-sponsored study presented at CROI 2024 showed semaglutide significantly reduces liver fat in people with HIV and MASLD [73], and (ii) the FDA approved a label expansion for Wegovy to include cardiovascular risk reduction based on long-term SELECT trial data [74]. These developments likely contributed to renewed optimism about semaglutide's broader therapeutic value.

However, the sentiment patterns among user groups vary. For example, we observe that organizational accounts and verified users have consistently more positive sentiment scores than individual accounts and unverified users. Users who are interested in Business exhibit moderate sentiment scores, reflecting the professional nature of corporate communications. Users interested in Politics exhibit the most negative and volatile sentiment, whereas the Beauty/Health category maintains the most positive sentiment among all groups throughout the timeline. Users with interest in Books maintained the most stable sentiment pattern. These patterns highlight how different user communities process health-related information through their respective contextual frameworks, with users valuing family and personal health showing the highest sentiment variation.

### ***The Impact of Semaglutide Shortage and Side Effect Reports***

In addition, we observe a significant downturn in sentiment across all user subpopulations during November 2022–January 2023. This universal decline aligns temporally with reports of semaglutide shortages [75-77] and reports adverse drug reactions [70], suggesting a public concern regarding the drug's availability and potential side effects. We apply an ITS regression to assess the impact of these shortages and determine whether the sharp decline in sentiment exceeds baseline trends, as displayed in Table 1. The baseline sentiment is significantly positive ($\beta_0$ = 0.172, $P$ = 0.003), with a modest but significant negative trend prior to the event ($\beta_1$ = −0.027, $P$ = 0.033). At the intervention point, an immediate and statistically significant drop occurs ($\beta_2$ = −0.430, $P$ = 0.004). While the post-intervention slope shows a slight positive trend ($\beta_3$ = 0.029, $P$ = 0.073), this effect does not reach conventional significance thresholds.

Figure 3 visualizes the ITS regression, displaying observed mean sentiment across bimonthly intervals. A vertical dashed line marks the November 2022 intervention, illustrating changes in level and slope. As shown in Figure 3, sentiment remains relatively stable and slightly positive throughout 2021 and most of 2022. A clear decline emerges in the November 2022–January 2023 period, after which sentiment consistently remains below zero, indicating a sustained downturn in public discourse. The ITS results suggest that the intervention period coincides with a significant immediate downturn in sentiment, followed by a slight, non-significant recovery trend thereafter.

**Table 1**. Interrupted Time Series (ITS) regression of average sentiment. The intervention point is set at November 2022–January 2023; Post Time indicates the slope change thereafter.

| Variable | β Coefficient | Std. Error | *P*-value | 95% CI |
|---|---|---|---|---|
| Constant | 0.172 | 0.047 | .003 | [0.070, 0.274] |
| Time | -0.027 | 0.011 | .033 | [-0.051, -0.003] |
| Event | -0.430 | 0.125 | .004 | [-0.700, 0.160] |
| Post Time | 0.029 | 0.015 | .073 | [-0.00, 0.06] |

**Figure 3.** Interrupted Time Series (ITS) analysis of average sentiment over time with external events. The dashed red line indicates the start of the intervention window.

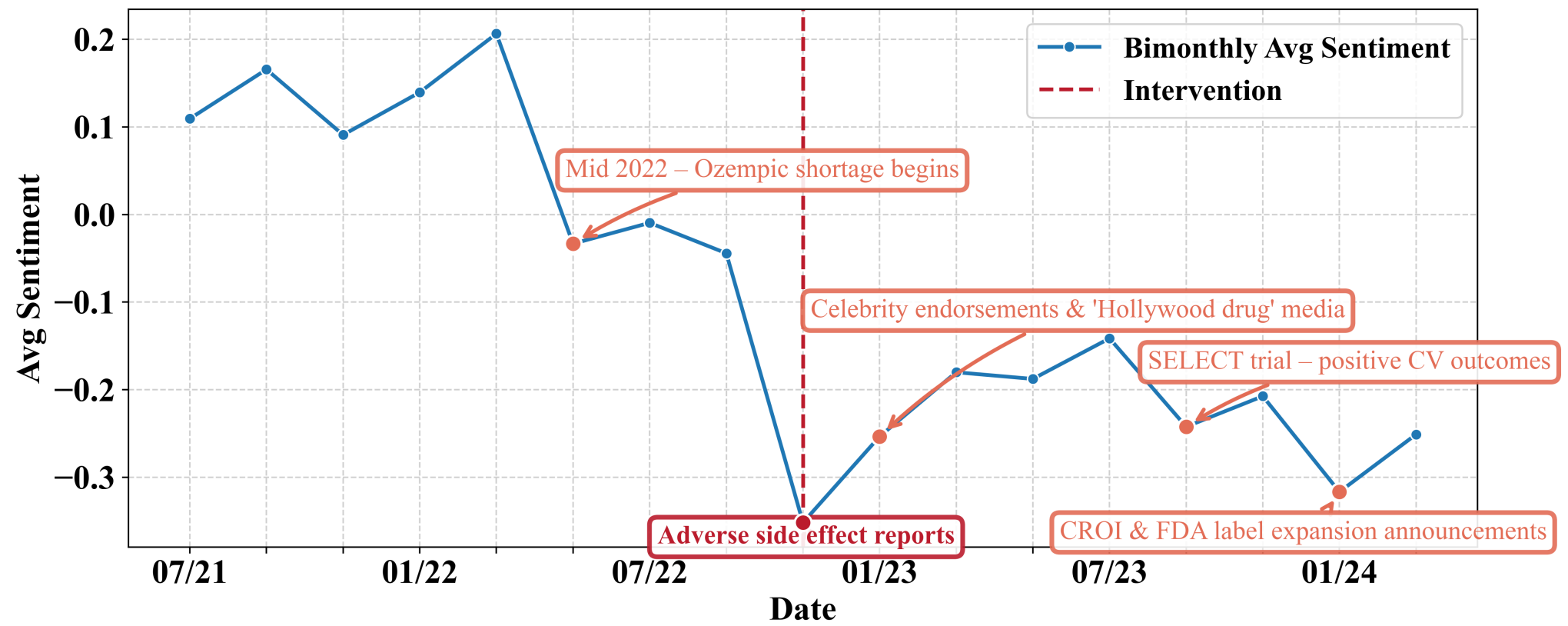

### *Sentiment Patterns across Subpopulations*

We observe that the per-user and per-post sentiment scores are similar; therefore, to avoid duplicated explanation and focus on user experience, we solely discuss per-user scores in this section (per-post scores and repost-excluded scores are available in Table S5 in Multimedia Appendix 1). The sentiment distribution among user subpopulations varies significantly, as illustrated in Figure 4. In addition, Table 2 presents 95% CIs for the estimated mean sentiment differences across key user groups. All comparisons yield statistically significant results ($P$ < .001), providing strong evidence that the observed differences are not due to random variation.

Based on Figure 4, male users exhibit more positive sentiment (averaged sentiment -0.20) than female users (averaged sentiment -0.32). Verified users, typically public figures or organizations with confirmed identities, expressed less negative sentiment (averaged sentiment -0.17) toward semaglutide than nonverified users (averaged sentiment -0.28). This contrast highlights how identity and accountability influence sentiment expression online. Organizational accounts expressed less negative sentiment compared to individual users. These findings suggest that organizations tend to frame their discussions about semaglutide in a more positive manner, possibly due to pharmaceutical marketing, professional communication standards, or endorsement practices. Within the United States, regional variations in sentiment toward semaglutide are evident in the analysis. Users in the Southeast region express most negativity (averaged sentiment -0.29) while users from the Northeast are less negative.

Lastly, Figure 5 presents the average sentiment scores for each combination of account type and user interest, measured per user and per post. Overall, individual users tend to express more negative sentiment than organizational accounts. For example, in the Travel category, individual sentiment is clearly negative (-0.24), while organizational sentiment is positive (0.12), yielding one of the largest absolute gaps between the two account types. This figure highlights that user sentiment varies widely across domains of interest. Health-related categories, particularly those tied to Beauty/Health & Fitness, generate higher level of positivity, while Business discussions are closer to neutral, reflecting a more corporate and less personal orientation. These findings underscore the importance of considering both account type and user interest when analyzing public sentiment.

**Table 2**. Estimated differences in sentiment across key subgroups, with 95% CIs.

| Comparison | Level | Mean 1 | Mean 2 | Mean Diff | 95% CI | $P$-value |
|---|---|---|---|---|---|---|
| Male vs Female | Per-post | -0.17 | -0.29 | 0.12 | [0.121, 0.129] | < .001 |
| | Per-user | -0.20 | -0.32 | 0.12 | [0.117, 0.128] | < .001 |
| Verified vs Nonverified | Per-post | -0.14 | -0.24 | 0.10 | [0.092, 0.102] | < .001 |
| | Per-user | -0.17 | -0.28 | 0.11 | [0.101, 0.116] | < .001 |
| Organizational vs Individual | Per-post | -0.01 | -0.24 | 0.23 | [0.222, 0.235] | < .001 |
| | Per-user | -0.04 | -0.28 | 0.24 | [0.232, 0.256] | < .001 |

**Figure 4**. Sentiment patterns across user subpopulations, measured per user and per post (the sentiment of a post can only be the value from -1, 0, and 1).

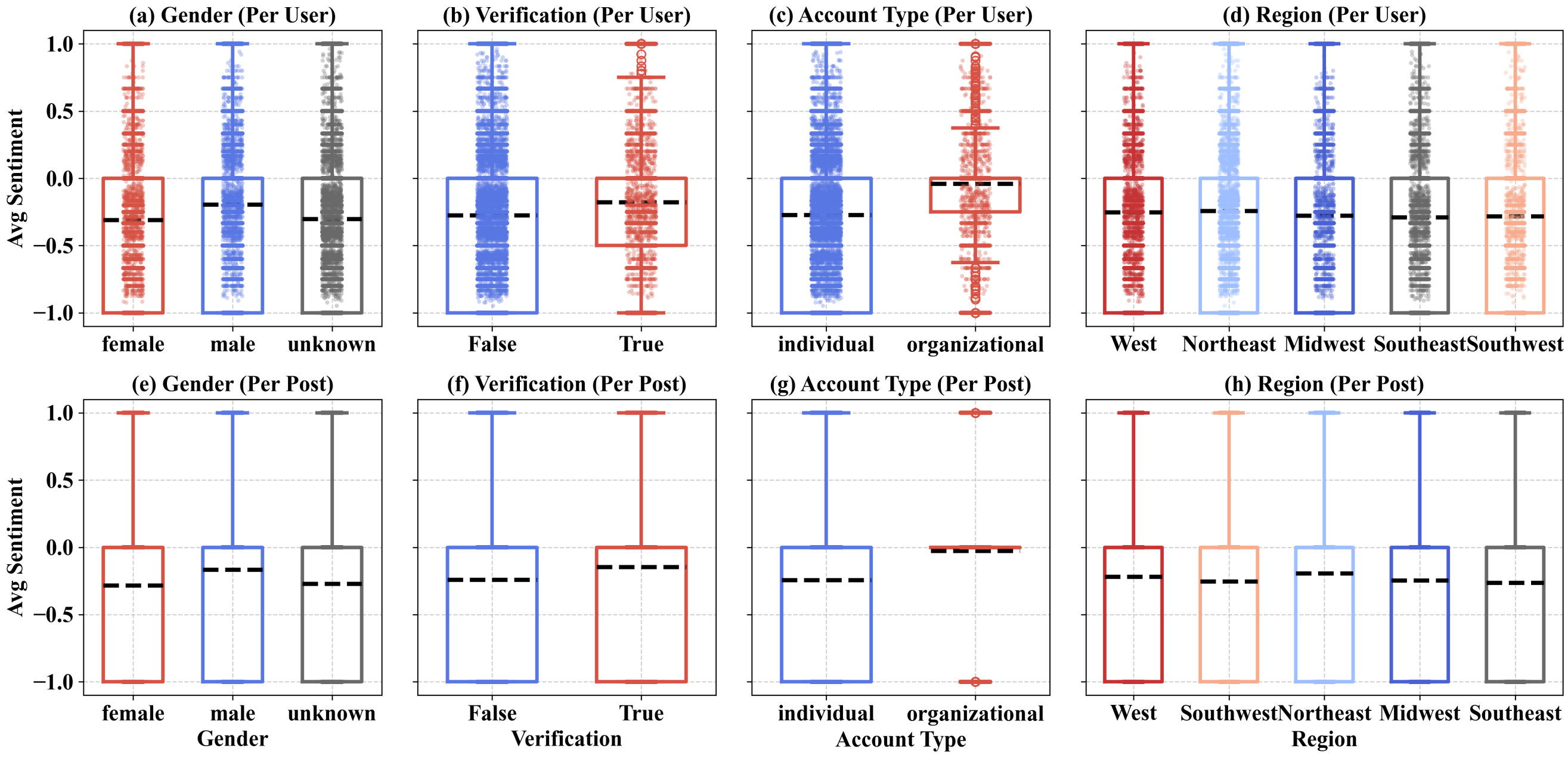

**Figure 5**. Combined heatmap of mean sentiment scores by account type and interest.

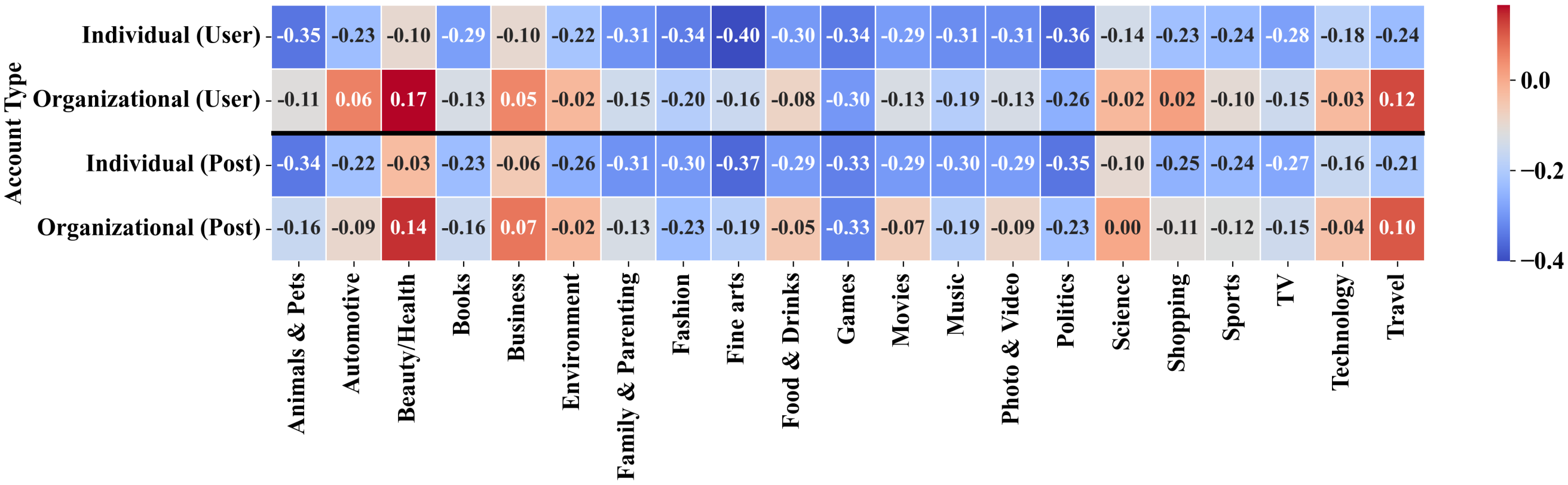

## RQ2 (Topic Results Across User Subpopulations)

For the topic analysis, we focus on original posts by successfully assigning umbrella topics to 207,849 out of 859,751 posts. Not all posts in the dataset receive topic assignments because we exclude duplicated reposts (to avoid counting the same content multiple times) and neutral-sentiment posts from the topic modeling process. We conduct topic modeling analysis by post rather than aggregating by user, which preserve the complete topic distribution for prolific users.

### *Topic Prevalence Across Subpopulations*

Figure 6 shows the prevalence of each topic within each user subpopulation, highlighting the variations in each topic's prevalence across subpopulations. The most common topic among subpopulations was T0 (Weight loss). However, the popularity of the other topics is less consistent across the subpopulations. We present a descriptive analysis of the topic prevalence results in the following paragraphs; confidence intervals and effect sizes (calculated via Cramér's V [78]) for these results are given in Table S7 in Multimedia Appendix 1.

T1 (Celebrities/politicians) is noticeably more popular among female users compared to male users, comprising 21% of posts from female users and 17.5% of posts from male users. Although there are more overall posts originating from male users, female users post more T1 posts. In addition, 13.4% of male user posts pertained to T4 (Drug authorities), compared to 9.1% of posts from female users.

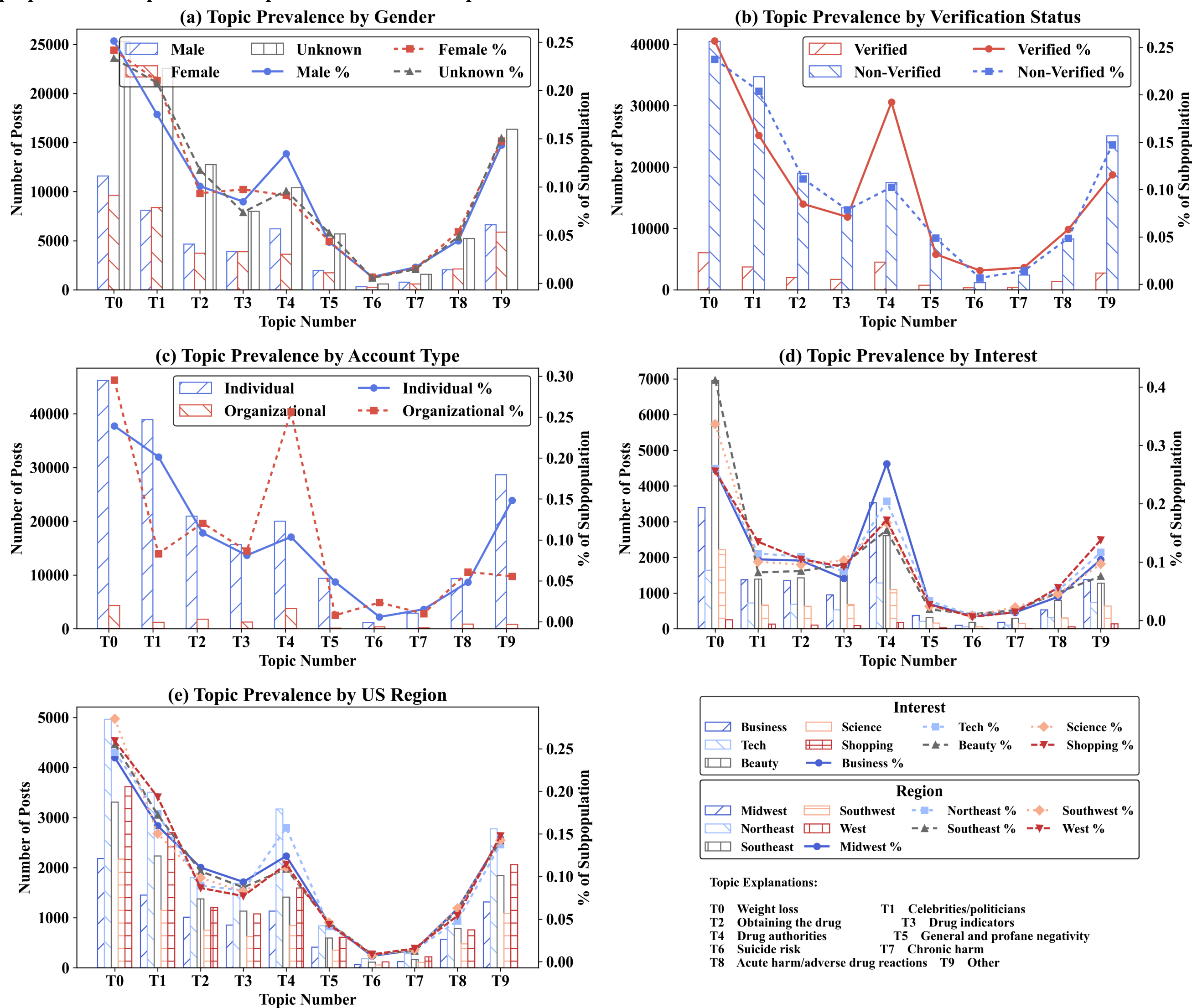

**Figure 6.** Prevalence of each topic within each user subpopulation. The bars and left axes measure the number of posts pertaining to each topic, and the lines and right axes measure the percentage of each subpopulation's posts that pertain to each topic.

**Table 3**. Number, name, and representative keywords for each of the ten umbrella topics.

| Topic No. | Topic Name | # of Posts | Example Representative Keywords |
|---|---|---|---|
| T0 | Weight loss | 50,542 | exercise, workout, eat, diet, appetite, skinny, obesity, craving |
| T1 | Celebrities/politicians | 40,120 | nikkifried, erikajayne, oliviawilde, rhianna, tuckercarlson, oprah, elonmusk, trump |
| T2 | Obtaining the drug | 22,724 | prescription, medication, walgreens, walmart, coupon, insurance, coverage, supply, ordered, appointment, affordable, shot, injection |
| T3 | Drug indicators | 16,948 | diabetes, inflammation, treatment |
| T4 | Drug authorities | 23,783 | pharma, novo, nordisk, doctor, physician, fda, goldman, economy, gdp, market |
| T5 | General and profane negativity | 9,488 | [swear words] |
| T6 | Suicide risk | 1,486 | suicide, autopsy, death, overdose |

| T7 | Chronic harm | 3,075 | addiction, cancer, tumor, alocepia, hair, hairline, dialysis |
|---|---|---|---|
| T8 | Acute harm/adverse drug reactions | 10,208 | nausea, diarrhea, constipation, pain, effect |
| T9 | Other | 29,475 | [anything that does not fit in with the other topics] |

Verified users are less likely to post profane posts, with T5 (General and profane negativity) comprising 3.2% of their posts, compared to 4.9% of posts from unverified users. However, verified users are about twice as likely to post about T6 (Suicide Risk) than unverified users. They are also more likely to create posts pertaining to T4 (Drug authorities); 19.2% of verified user posts belong to T4, compared to 10.3% from unverified users.

Examining the most prevalent topics among individual accounts and organizational accounts, the most striking difference is the very low number of profane posts within the organizational account subpopulation. T5 comprises just 0.8% of organizational posts, compared to 4.9% of posts from individual users. As companies and organizations likely do not want to damage their reputation by posting profane content, this result is in line with our expectations. On the other hand, organizations are about four times more likely than individuals to create posts pertaining to T6 (Suicide Risk), perhaps due to medical organizations posting warnings about potential side effects of semaglutide. T4 is far more common among organizational accounts, with 25.7% of organizational posts belonging to T4 compared to 10.4% from individual users.

Dividing the users by interest reveals several differences in topics of discussion. Notably, the subpopulation consisting of users interested in Business is the only subpopulation in which T0 is not the most prevalent topic. T4 is the most common topic among users interested in Business. Users in the Business subpopulation appear to be more interested in the economic impact of semaglutide, as opposed to its usage in weight loss treatment. As expected, T0 is by far the most popular topic among users interested in Beauty/Health. Lastly, the prevalence of each topic is mostly consistent across different U.S. geographic regions. However, there are some variations; for example, T4 is noticeably more popular in the Northeast compared to other regions.

### *Topic Prevalence Over Time*

Figure 7 shows the number of posts pertaining to each umbrella topic posted during each bimonthly period from July 2021 to April 2024. To assess the evolution of topic prevalence over time, we first present an exploratory analysis of external events that may have influenced the topic trends. All topics rose in popularity from July 2021 to April 2024. This result is consistent with the general increase in popularity of semaglutide. T3 surged in popularity during November 2022–January 2023. This trend aligns with the initial FDA approval of Wegovy for adolescents [69], which occurred on December 23, 2022. During September 2023–November 2023, T0 (Weight loss) increases in prevalence, while T8 (Acute harm/adverse drug reactions) and T6 (Suicide risk) decline. These changes align with the release of positive cardiovascular outcomes for Wegovy from the SELECT trial [72].

**Figure 7**. The number of posts pertaining to each topic over time.

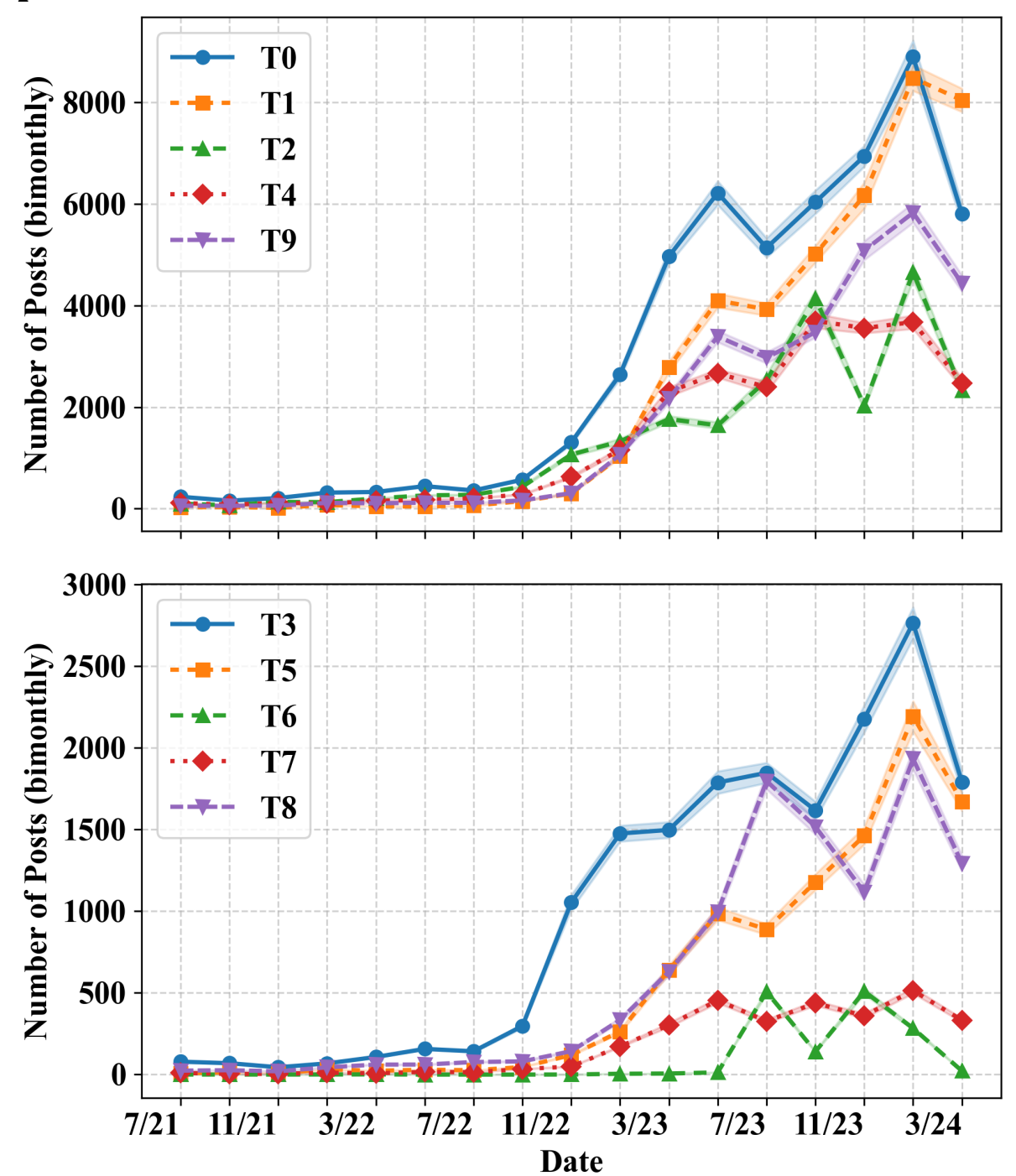

Starting in July 2023–September 2023, there is a noticeable uptick in the number of posts pertaining to T6 (Suicide risk) and T8 (Acute harm/adverse drug reactions). To determine whether this change exceeds baseline trends, we apply an ITS regression, The ITS results are visualized in Figure 8 (full results are available in Multimedia Appendix 1 in Table S6). A statistically significant ($\beta_2$ = 449.51, $P$ < .001) increase occurs at the intervention point, followed by a significant downward trend ($\beta_3$ = -83.10, $P$ = .02). The ITS results suggest that the intervention window coincides with a sharp increase in the prevalence of T6. The surge in the popularity of T6 may have been caused by a statement released by the European Medicines Agency on July 11, 2023, acknowledging "*about 150 reports of possible cases of self-injury and suicidal thoughts*" from "*people using liraglutide and semaglutide medicines*" [79].

**Figure 8**. ITS analysis of T6 (Suicide Risk). The intervention window is July 2023-September 2023.

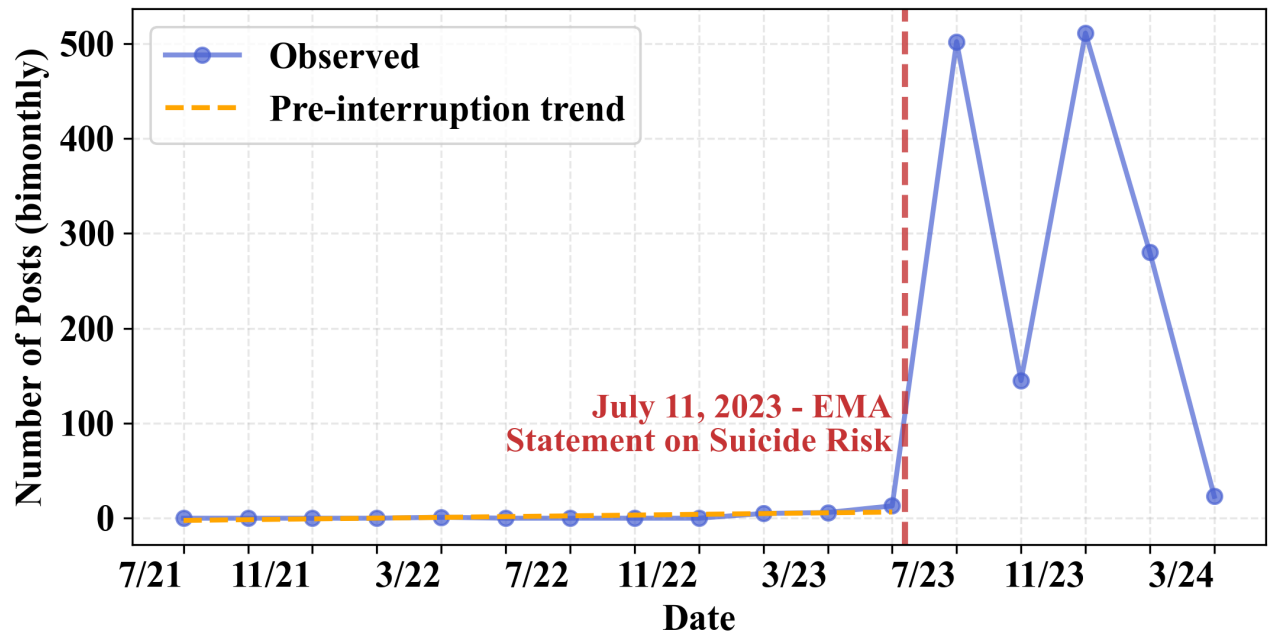

### *Breakdown of Topics by Sentiment*

Figure 9 shows the breakdown of sentiment scores within each topic, within the gender, verification status, and account type subpopulations. This figure connects our sentiment and topic analyses and provides insights into potential sources of user negativity and positivity. For example, topics T5 (General and profane negativity), T6 (Suicide risk), T7 (Chronic harm), and T8 (Acute harm/adverse drug reactions) are almost entirely negative.

The breakdown also reveals notable differences in sentiment within topics. In T4, for instance, male-identified accounts contribute a significant volume of positive posts, whereas female-identified accounts are almost exclusively negative (Figure 9a). A similar, though less pronounced, pattern is visible in T9, where male accounts again show a larger share of positive sentiment. Furthermore, in T0, organizational accounts exhibit a much higher *proportion* of positive sentiment compared to individual accounts, whose posts in that topic are predominantly negative (Figure 9c).

**Figure 9.** Breakdown of sentiment scores by topic number, within various user subpopulations.

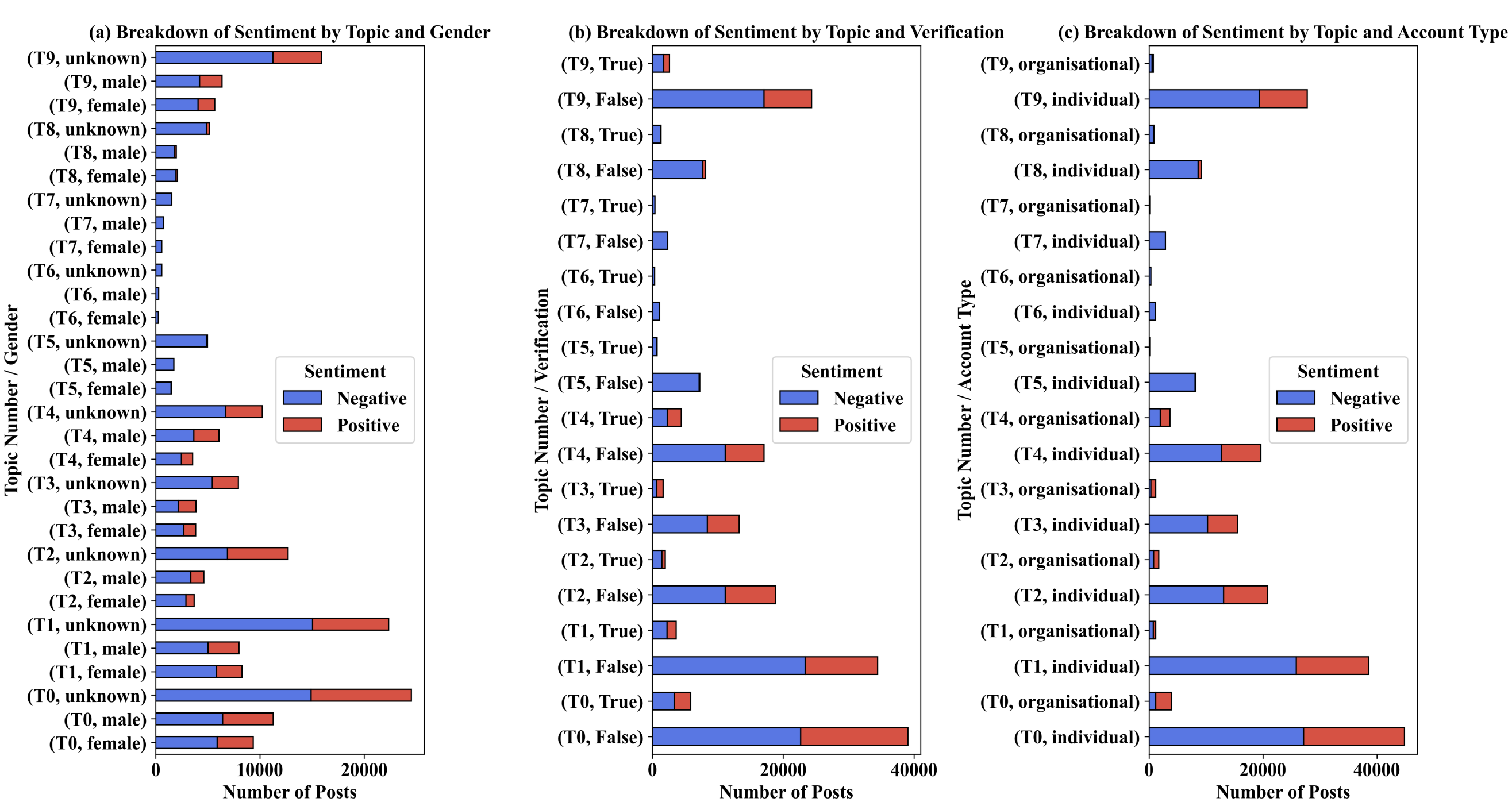

## Discussion

### Key Findings

The primary takeaway from our results is that, although semaglutide has generated considerable attention, its reception from individual X users is generally negative. This negativity suggests an overall public skepticism regarding the drug's efficacy, accessibility, and potential side effects, which are consistently highlighted in user discussions. Such negative sentiment is particularly notable in individual accounts as compared to the positive or neutral sentiment found more often in organizational accounts.

The temporal analysis demonstrates a general decrease in sentiment over time. In addition, we observe a notable shift in sentiment during late 2022, when regulatory announcements related to adverse effects and safety warnings surfaced [70]. This period sees a sharp increase in negative sentiment, which aligns with concerns raised about the drug's safety. The general throughline of negativity beginning in mid-2022 aligns temporally with FDA confirmation that Wegovy (semaglutide) is in shortage as well as heightened media coverage of semaglutide's off-label use for weight loss. The observed negativity likely reflects public concern about limited access and fairness, as shortages reported by the FDA were simultaneously amplified in professional and mainstream outlets. These shortages and their subsequent resolution are noted in multiple sources, including the FDA Drug Shortages Database and trade publications documenting semaglutide's removal from the shortage list and its ongoing legal and ethical implications [76, 77].

In a similar note, discussions concerning serious impact spike following the European Medicines Agency's report on suicides on July 11, 2023 [79]. These findings underscore the impact that regulatory decisions and public health announcements can have on shaping public perceptions, particularly when safety and efficacy concerns are at the forefront of the discussion. The ITS analyses reinforce our interpretation that the drop in sentiment and uptick in T6 exceed baseline trends, likely having been influenced by external events. Given the temporal alignment and the magnitude of the changes, it is likely that public sentiment is shaped in response to reports of adverse effects and that the EMA statement on self-harm reports influenced public discussions.

In addition, user subpopulations show variance in sentiment and topic discussion. Users who are interested in Beauty/Health has the most positive sentiment and the highest prevalence of T0 (Weight loss). Male users are slightly more positive than female users, though they appear less interested in T1 (Celebrities/politicians). The observed gender differences may be influenced by the increased media attention surrounding celebrities endorsing the drug, which is more prominently featured in female-driven narratives about weight loss and beauty [80, 81]. On the whole, verified users and organizations are more positive than their counterparts. Additionally, T4 (Drug authorities) is more prevalent among verified users and organizations, while T1 (Celebrities/politicians) and T5 (General and profane negativity) are more prevalent among unverified users. These differences suggest that verified users (often public figures or organizations) tend to use a more conservative tone. On the other hand, nonverified users (often individuals) tend to emphasize personal concerns, particularly regarding side effects and affordability. This difference highlights how the identity and motivations of the speaker can influence sentiments, with verified accounts potentially downplaying issues for commercial or reputational reasons, while individual users are more candid about their negative experiences.

### Practical Implications

Our findings have several practical implications for stakeholders, particularly healthcare providers and pharmaceutical companies. Overall, our results underscore the need for transparent communication strategies. These strategies should prioritize addressing concerns raised by users, including issues related to accessibility, side effects, and the drug's overall safety. Clear communication can help bridge the gap between public perception and medical realities, while fostering trustworthy and informed decision-making.

For pharmaceutical companies, transparency in messaging about the use and effects of semaglutide is critical. Our observations highlight a gap in sentiment between individual and organizational accounts. One potential contributing factor is the societal emphasis on beauty and weight loss, which can be amplified by advertising and endorsements from influential figures [6]. This phenomenon often skews public perception and drives expectations.

Given that regulatory announcements and safety warnings significantly shape public sentiment [79], pharmaceutical companies should adopt proactive approaches to build credibility. This includes consistent and timely communication that addresses misconceptions and reinforces the drug's benefits and limitations.

### Limitations

While this study offers comprehensive insights into public perceptions of semaglutide on X, some limitations must be acknowledged. First, our dataset is limited to English-language posts, and our regional sentiment analysis is limited to the United States. While this constraint is necessary for linguistic consistency in sentiment and topic modeling, it may exclude important perspectives from non-English-speaking users, introducing a potential source of bias. Second, gender classification in our dataset is limited to three categories: male, female, and unknown. We recognize that gender is not binary and includes a spectrum of identities. However, the Brandwatch data source only provides these limited groupings. As a result, our analysis may not capture the experiences of gender-diverse users, representing a gap in inclusivity. Third, our study only makes use of data from X platform, potentially limiting the generalizability of the results to different platforms (e.g., Reddit and clinical forums). Future work may study additional platforms to build a broader understanding of user experience. Lastly, although our work identifies numerous trends in public attitudes towards semaglutide, we do not interpret it as a causal impact of real-world reports (e.g., semaglutide usage, adverse events) on online discourse.

## Conclusions

Public interest in semaglutide has greatly increased in recent years. This study explores X users' experiences with semaglutide via an analysis of 859,751 X posts created between July 2021 and April 2024. We observe a general decrease in sentiment across most user subpopulations over time, with a particularly noteworthy decrease occurring in November 2022. Our research highlights the complex dynamics of user experiences with semaglutide, driven by a combination of user demographics, regional factors, and external events such as regulatory announcements. The practical implications of these findings are crucial for healthcare communicators and pharmaceutical companies seeking to engage with the public in a more informed, responsive, and regionally targeted manner. Future research should focus on further unraveling the role of side effects in shaping public opinion and exploring how sentiment changes in response to evolving health-related information.

## Data Availability

Information about accessing an anonymized version of the dataset generated and analyzed during this study is available in Multimedia Appendix 1.

## Authors' Contributions

Conceptualization: LL, JL
Data curation: LL, PM, GL
Formal analysis: PM, GL
Investigation: PM, GL
Methodology: PM, GL, LL
Project administration: LL, JL
Software: PM, GL, LL
Supervision: LL, JL
Visualization: PM, GL, LL
Writing – original draft: PM, GL, LL
Writing – review & editing: PM, GL, LL

## Conflicts of Interest

None declared.

## Multimedia Appendix 1

Additional details about the user subpopulations, dataset availability, elbow method results, umbrella topics, umbrella topic stability checks, repost-excluding sentiment sensitivity analysis, and topic modeling results can be found in Multimedia Appendix 1.

## Multimedia Appendix 2

The complete mapping of all 200 initial clusters to the 10 umbrella topics is available in Multimedia Appendix 2.